\pdfoutput=1

\documentclass[1p,times,11pt]{article}
\usepackage{graphicx}

\usepackage{amssymb}
\usepackage{mathrsfs}
\usepackage{latexsym}
\usepackage{amsmath}
\usepackage{subfigure}

\usepackage[figuresright]{rotating}

\begin{document}


\title{Some topics in the kinetics of protein aggregation}
\author{\textsc{J. D. Gunton, Wei Li, Ya Liu, Toni Perez}\\
Department of Physics, Lehigh University, Bethlehem, PA 18015\\ \\
\textsc{S. J. Khan, A. Chakrabarti}\\Department of Physics, Kansas State University, Manhattan, KS 66503}





\maketitle

\begin{abstract}
Preliminary results are presented for the kinetics of phase separation in three distinct models of protein aggregation.
The first is a model of the formation of spherical microcrystals of insulin via an initial formation of fractal clusters of insulin.
The results of our Brownian dynamics study of this model are in qualitative agreement with a recent experimental study \cite{Bromberg_05_01} of microcrystal formation
from aqueous mixtures of insulin.  A second work involves a theory for the formation of metastable bundles of sickle hemoglobin from fibers, based on a recent generic theory of bundle formation \cite{Grason_07_01}.  We also discuss a model for the microscopic formation of these fibers.  Finally, we discuss preliminary results for
the kinetics of cluster formation for a six patch model of protein crystallization.\\ \\
Keywords: protein aggregation, brownian dynamics, sickle hemoglobin, patch model
\end{abstract}






\section{Introduction}
In the spirit of this Workshop, we present some preliminary results for several problems that involve the kinetics of protein aggregation.  This subject is an active field of research that includes studies of protein crystallization as well as certain biomedical problems such as human cataract formation, sickle cell anemia, Alzheimer's disease and various problems in drug delivery.  The specific problems discussed below have in common that they involve the kinetics of phase separating systems.  A review of recent developments in this field is contained in \cite{Gunton_07_01}.

\section{A Model of Microcrystal Formation in Insulin Solutions}
A standard method of preparing microcrystals of insulin for use in drug deliver is to precipitate insulin from aqueous solutions using
zinc salts.  This is the same technique as is used in many kinds of protein crystallization.  However, in 2003 Bromberg et al \cite{Bromberg_05_01} used an alternative method in which the aqueous solution was prepared near the isoelectric point of insulin, in order to minimize Coulomb interactions.  Rather than using a salt
to precipitate insulin from solution, they used polyethylene glycol (PEG), which is known to induce attractive interactions between biomolecules via a depletion attraction.  The authors first quenched the system to a low temperature and found that the insulin particles precipitated from solution in the form of a fractal network.  The fractal nature was established  by continuous-angle small angle light scattering, which showed a fractal dimension of 1.8.  This dimension is characteristic of the case of
diffusion-limited cluster-cluster aggregation.  They then stirred and subsequently diluted the sample to break up the fractal network into a relatively monodisperse set of microcrystals.  This promises to be an alternative method of microcrystal preparation in the drug delivery of insulin.\\

Our group has recently carried out a Brownian dynamics simulation of a model of this experiment, to see if we are able to capture the qualitative features of the experiment.  Our Hamiltonian consists of a hard core interaction between the insulin molecules, together with a short-range attractive interaction given by the Asakura-Oosawa model of depletion attraction.  The latter is a semi-quantitative approximation of the depletion interaction induced by the PEG polymers.
In our Brownian dynamics (BD) simulations
, we consider a three-dimensional system
of size $ L = 64 \sigma$ in units of the insulin molecular diameter $\sigma$. All other length scales are measured in units of $\sigma$ as well. We consider the case of low volume fraction $f=0.02$
for a system of 10,013 insulin molecules.
The equations of motion for the BD simulation are
\begin{equation}
\ddot{\vec{r}}_{i} = - \nabla  U_{i} -\Gamma \dot{\vec{r}}_{i} +\vec{W}_{i}(t)
\end{equation}

\noindent
where $\Gamma$ is the friction coefficient and $\vec{W}_{i}$, the random force acting on each insulin particle i, is
a Gaussian white noise satisfying a fluctuation-dissipation relation. Hydrodynamic interactions,
including lubrication forces are ignored in the simulation. The potential U acting
upon each insulin monomer has a twofold contribution: the two-body depletion potential of the
Asakura-Oosawa-Vrij ($U_{AO}$) plus a repulsive hard-core-like interaction ($U_{HC}$) given by the
following expressions:
\begin{equation}
U(r_{ij}) = U_{AO}(r_{ij}) + U_{HC}(r_{ij})
\end{equation}
where
\begin{equation}
\frac{U_{AO}(r_{ij})}{k_{B}T}= \phi_p (\frac{1+\xi}{\xi})^3[\frac{3r_{ij}}{2(1+\xi)}-\frac{1}{2}(\frac{r_{ij}}{1+\xi})^3 - 1],      r_{ij} < (1+\xi),
\end{equation}
and is zero for $r_{ij} > (1+\xi)$.  The hard core potential is given by
\begin{equation}
\frac{U_{HC}(r_{ij})}{k_{B}T} = r_{ij}^{-\alpha}.
\end{equation}

In Eq. 3, $\xi$ is the size-ratio between a polymer coil and a colloidal particle which controls the
range of the depletion interaction in the Asakura-Oosawa-Vrij model and $\phi_p$ is the polymer
volume fraction which controls the strength of the interaction.  Our simulations are for
$\xi = 0.1$.  In the hardcore-like repulsive interaction given by Eq. 4, we have set $\alpha= 36$. Values of $\alpha < 36$ have been reported
to lead to anomalies when a mimic of the hard-core potential is required in the potential [18,33]. The total pair potential
$U=U_{AO}+U_{HC}$  has a minimum value ($U_m$) that depends on $\xi$ and $\phi_p$. In
what follows, we will often characterize the strength of the potential in terms of the absolute
value of the minimum potential depth, $U_m=|U_{min}|$. We choose $\Gamma = 0.5$ and a time step
$\Delta t = 0.005$ in reduced time units of $\sigma (m/k_{B}T)^{1/2}$, with $m=1$.  For this choice of $\Gamma$, particle motion
is diffusive for $t \gg \frac{1}{\Gamma}$,
i.e.$ t \gg 2$ in our units. Periodic boundary conditions are used
to minimize wall effects. All simulations start from a random initial monomer conformation and
the results for the kinetics are averaged over several runs.

\subsection{Results}
\begin{figure} \center
\includegraphics[width=8cm]{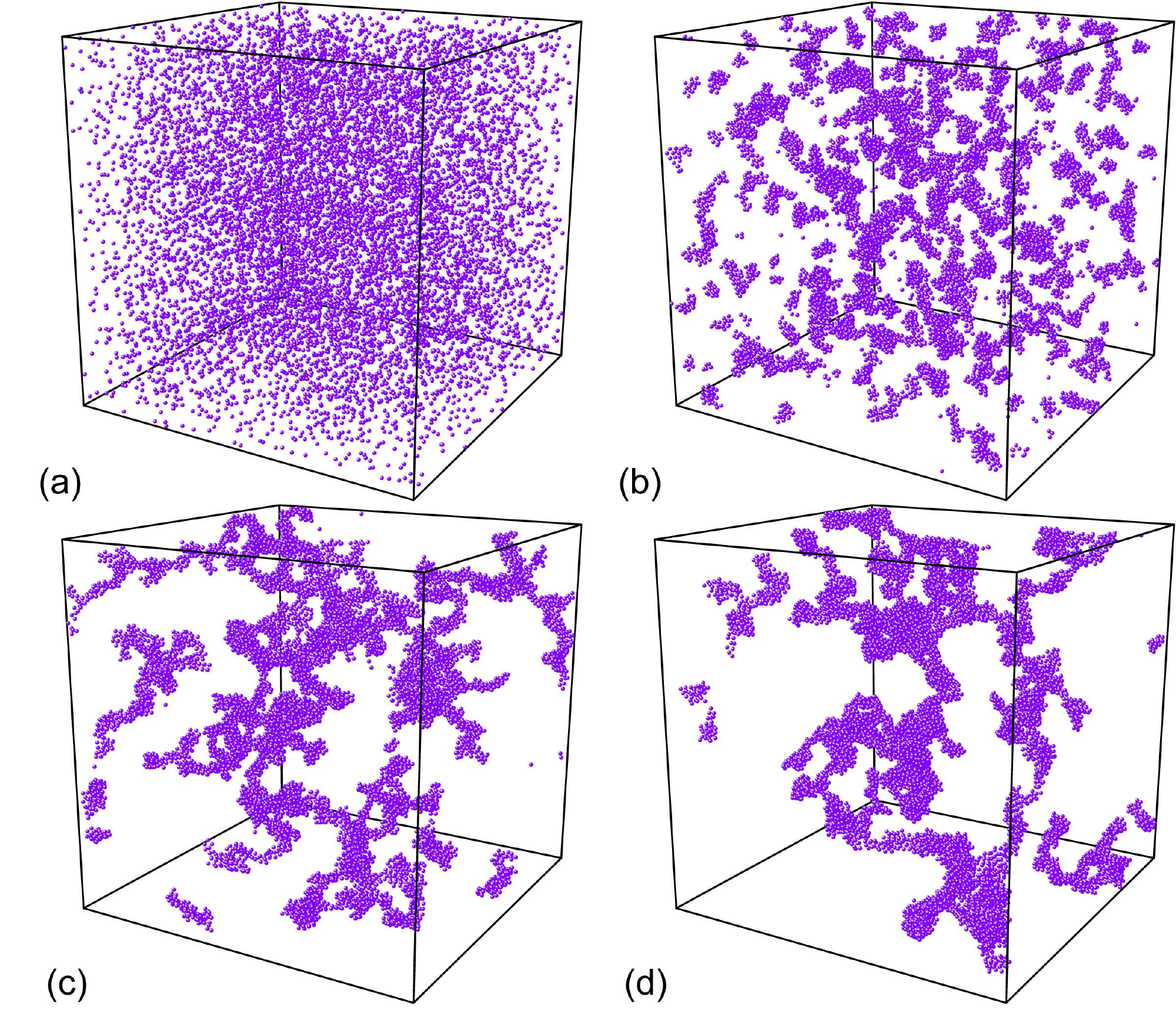}
\caption{Morphology of cluster formation ($\xi=0.1$, $f=0.02$) for a deep quench ($|U_m|=10.0k_{B}T$)  into the
two phase gas-solid region at various times: (a) $t=0$, (b) $t=50$, (c) $t=1,000$, and (d) $t=10,000$.}
 \label{fig_FIG1}
\end{figure}

\begin{figure} \center
\includegraphics[width=12cm]{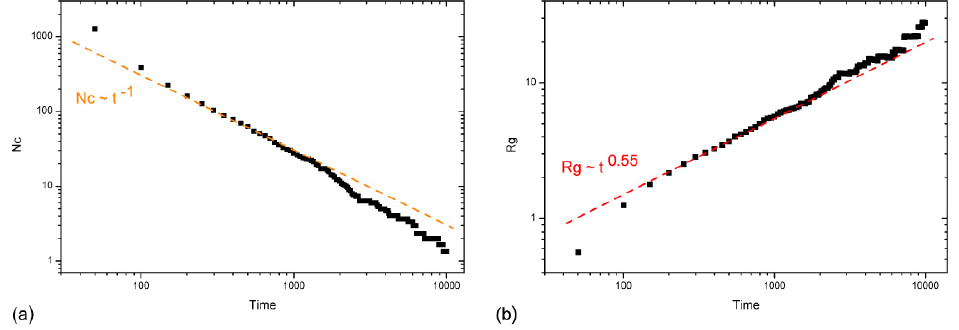}
\caption{Plot of (a) number of clusters,  $N_c$,  and (b) radius of gyration, $R_g$,  as a function of time in a
log-log plot, for a $10k_{B}T$, $f=0.02$, $\xi=0.1$ system; this shows slopes of $-1$ and $0.55$, respectively, in the
early stages, in agreement with DLCA.}
 \label{fig_FIG2}
\end{figure}

\begin{figure} \center
\includegraphics[width=12cm]{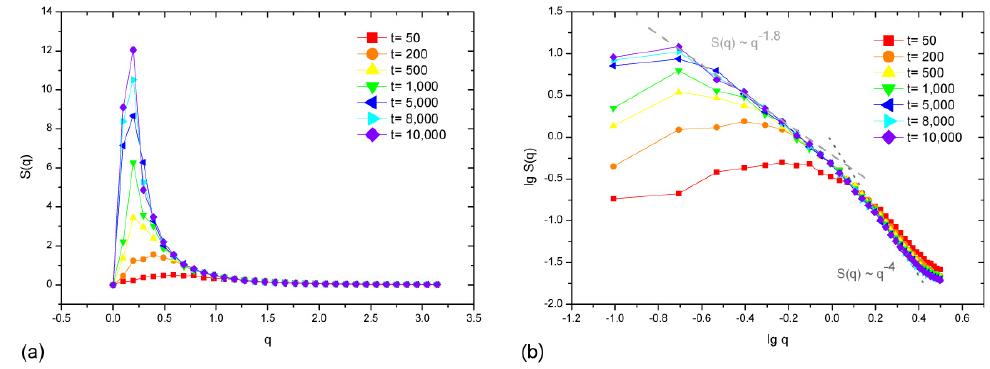}
\caption{(a) Structure factors and (b) log-log plot of structure factors at several times for
$|U_m|=10.0k_{B}T$. Dashed line indicates fractal clusters with $S(q)\sim q^{-D_f}$, where $D_f=1.8$. The dotted line
indicates the Porod regime $S(q)\sim q^{-(d+1)}$, $d=3$.Thus the clusters are hybrid fractals with short-range crystalline order and long-range fractal morphology.}
 \label{fig_FIG3}
\end{figure}

\begin{figure} \center
\includegraphics[width=8cm]{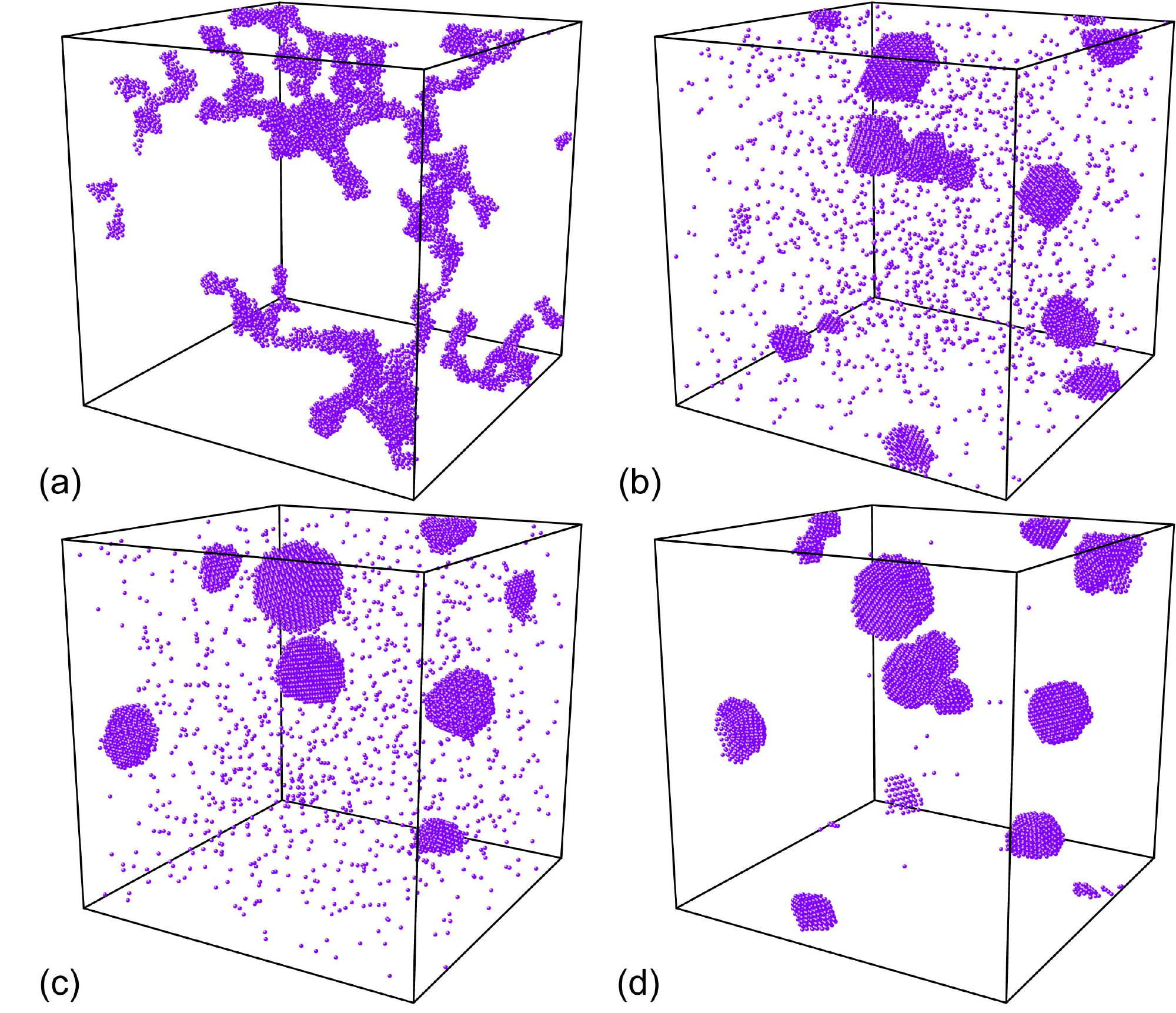}
\caption{(a) Largest cluster from a deep quench $|U_m|=10.0k_{B}T$, $\xi=0.1$, $t=10,000$. Cluster
morphology at $t=5,000$ for break-up of the fractal clusters and formation of spherical aggregates.  (b)
Heating up the entire system  (a) to $|U_m|=2.88k_{B}T$ for $t=5,000$; (c) keeping the entire system
in (b) at $|U_m|=2.88k_{B}T$ for another $t=5,000$ and (d) cooling the entire system in (b) to
$|U_m|=3.72k_{B}T$ for $t=5,000$.}
 \label{fig_FIG4}
\end{figure}

We summarize here some of the results of our simulation.  We first quench the system deep into the two phase gas-solid region (with $|U_m|=10.0k_{B}T$)
and study the kinetics of the resultant cluster formation.  Figure \ref{fig_FIG1} shows the morphology of the system for various time steps.  It appears that
the system initially phase separates through the formation of fractal clusters.  We have verified this in more detail in various ways; Figure \ref{fig_FIG2}
shows the behavior of the number of clusters, $N_c$ and the radius of gyration, $R_g$ as a function of time, in a log-log plot.  These have slopes of $-1$ and $0.55$, respectively, in the early stages of development, in agreement with the diffusion limited cluster-cluster aggregation theory (DLCA).  We have also studied the structure factor of the system at various times (see Figure \ref{fig_FIG3}) and found that this behaves like $S(q)\sim q^{-D_f}$, where $D_f=1.8$, for small q, consistent with fractals with a dimension $D_f=1.8$.  
The large q behavior is consistent with Porod's law, which characterizes the scattering from compact clusters, namely $S(q)\sim q^{-(d+1)}$, $d=3$.  Thus the clusters are hybrid fractals with short-range crystalline order and long-range fractal morphology. To simulate the experimental situation in which the insulin mixture is stirred and diluted after the formation of fractals, we have taken out the largest cluster, put it into another simulation box and "heated" the system to $|U_m|=2.88k_{B}T$.  Its subsequent time evolution is shown in Figure \ref{fig_FIG4}.  As can be seen this process results in a break-up of the fractal cluster into spherical aggregates, as shown, say, in panel c) of that figure.  
We then "cooled" the system in b) to $|U_m|=3.72k_{B}T$ to further stabilize this distribution of droplets.  By this process, we have been able to show that we can reproduce the essential features of the experimental study by Bromberg et al \cite{Bromberg_05_01} via the Asakura-Oosawa model for depletion attractions induced by PEG.  
A more quantitative theory of their experiment would require the inclusion of additional forces that would lead to a temperature dependent behavior of the system, as seen experimentally.

\section{A Theory for the Finite Bundle Size of Sickle Hemoglobin Molecules}
Aggregates, or bundles,  of twisted protein fibers, such as sickle hemoglobin and actin, are important examples of biopolymers in which elastic interactions play a crucial role in determining the (metastable) bundle radii.  In one recent paper \cite{Ferrone_03_01} Turner et al.  proposed a model for stabilizing approximately 20 nm diameter bundles of sickle hemoglobin (HbS) fibers.  They constructed the free energy per unit volume, G, needed to create a fiber bundle, where $G=F-\frac{\psi}{\Lambda}$, using continuum elasticity theory. Here F is the distortion free energy per unit volume of a bundle of radius R and pitch length $\Lambda$ and $\psi$ is a positive Lagrange multiplier that controls the pitch length.  From G they predicted the physical properties of the fiber bundle, such as the equilibrium (metastable) bundle radius $R_c$, where in the latter case they minimized G with respect to R.  However, we believe their analysis is incorrect for two reasons, the first being the use of the free energy density, G, rather than the total free energy $R^2LG$, to determine $R_c$.  The second is their omission of the binding energy between fibers, which in classical nucleation theory of spherical droplets corresponds to the driving
force for nucleation. We present a corrected version of their analysis below. Our approach is the same as that of Grason and Bruinsma \cite{Grason_07_01}, who determined the critical bundle size for aggregates of filamentous actin. \\
According to classical homogeneous nucleation theory \cite{Grason_07_01,Oxtoby1992,Onuki_02_01}, the critical "droplet" size  corresponds to the maximum of the total free energy $R^2LG$, which is significantly different from the maximum of the free energy density $G$. A simple example is the nucleation of a spherical droplet \cite{Onuki_02_01}, in which the total free energy is given by $ 4\pi R^2 \gamma - \frac{4\pi}{3}R^3\epsilon$, where $\gamma$ is the surface tension and $\epsilon$ is the free energy density difference between the metastable and stable phases.  $G(R) = \frac{3 \gamma}{R } - \sigma$. The critical radius $R_c$ is determined by maximizing this total free energy (not the free energy per unit volume) with respect to R.  The analogous argument for the heterogeneous nucleation of the fiber bundle involves calculating the total free energy involved in creating this bundle from an aggregate of fibers of (fixed) length L.  

\begin{figure}[htbp]
\centering
\includegraphics [width= 10cm]{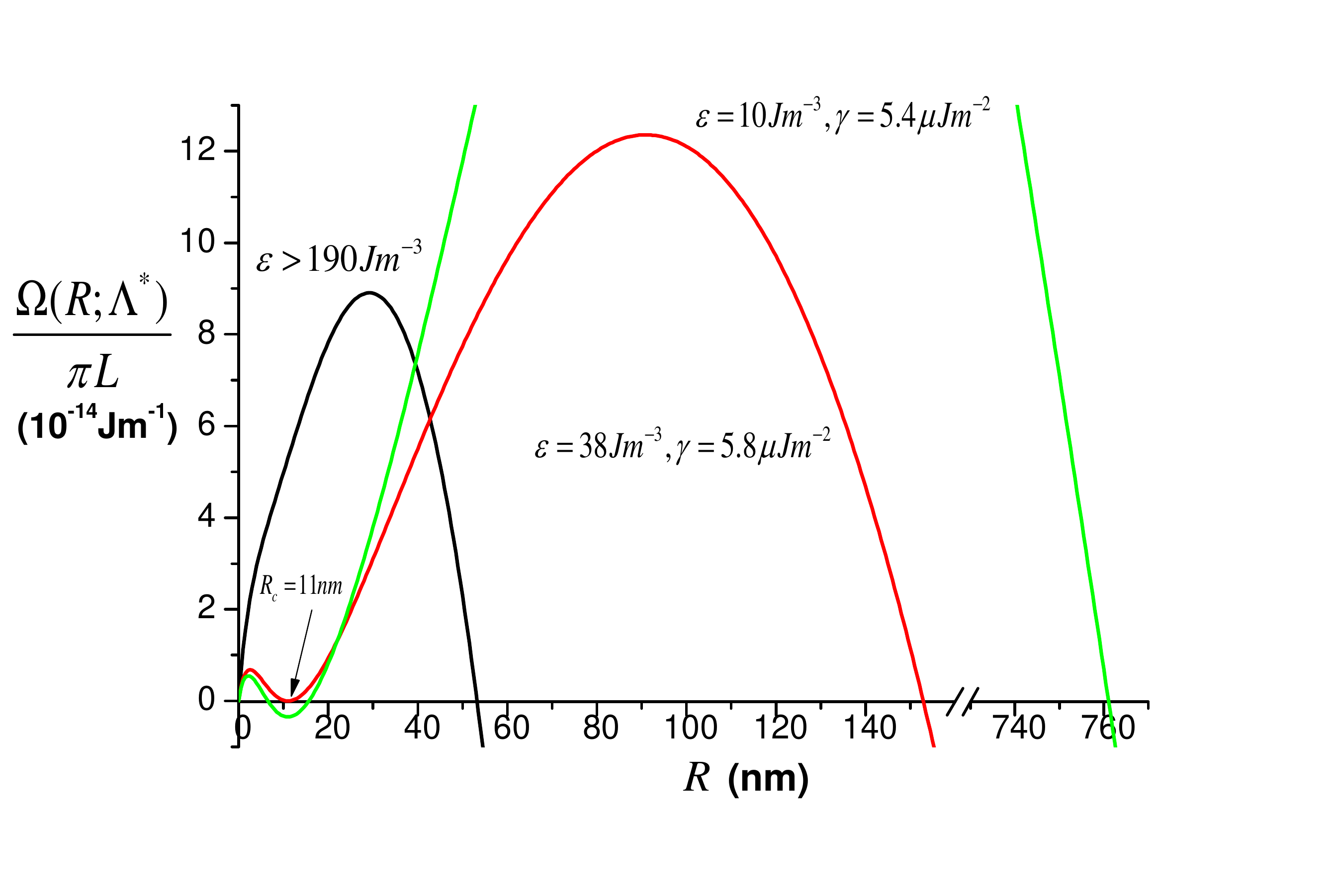}
\caption{Plot of the free energy $\Omega(R;\Lambda^{*})$ per unit length as a function of the fiber radius, R, using the experimental values for HbS given in the text for
 $\epsilon > 190 Jm^{-3}$  (black),  $\epsilon =  38Jm^{-3}$ (red), and  $\epsilon =  10Jm^{-3}$ (green).  One local minimum occurs at $R = 11 nm$ which corresponds to the (metastable) equilibrium radius of HbS when $ \epsilon = 38 Jm^{-3}, \gamma = 5.8  \mu Jm^{-2}$.}
\label{Fig:HbSEnergy}
\end{figure}

The grand potential of a twisted fiber $\Phi_{G} $ as a function of pitch $\Lambda$ and radius R, includes the contributions from the surface tension, extension or compression, bending, twisting, binding and chemical potential:
\begin{eqnarray}
\Phi_{G}  &=& \Omega - \mu \frac{R^2}{a^2} \nonumber \\
\Omega &=&  \pi L (2\gamma R +  ER^2\frac{\frac{\pi^4 a^2R^2}{16} + \frac{\pi^4R^4}{96}}{\Lambda^4} - \frac{R^2\psi}{\Lambda} -  R^2 \epsilon) 
\label{Eq:FreeEnergy}
\end{eqnarray}
where L is the fiber length, E the extensional modulus, a the radius of a protofilament, and $\mu$ the chemical potential of protofilaments. $\psi$ is related to the twisting stiffness \cite{Grason_07_01,Ferrone_03_01}.  Equation \ref{Eq:FreeEnergy} contains an additional term $-R^2\epsilon $ due to the aggregation energy \cite{Oxtoby1992,Onuki_02_01} between fibers that is not present in Turner et al. \cite{Ferrone_03_01}.  In the limit of $L \rightarrow \infty$, the grand potential is dominated by $\Omega$. 
Therefore the equilibrium pitch is determined by $\frac{\partial \Omega}{\partial (\pi L \Lambda)} |_{\Lambda = \Lambda^*}= 0$, which reduces $\Omega$ to

\begin{eqnarray}
\frac{\Omega(R;\Lambda^{*})}{\pi L} =  2\gamma R -  \frac{3^{4/3}\psi^{4/3}}{2\pi^{4/3}E^{1/3}} \frac{R^{4/3}}{(6a^2 + R^2)^{1/3}} -  R^2 
\label{Eq:SimEnergy}
\end{eqnarray}

Using experimental values for HbS  of $a = 4 nm$, $E = 51 MPa$, $\psi = 3.5 \times 10^{-4} Jm^{-2}$ \cite{Ferrone_03_01,Jones2003}, we find that $\Omega(R;\Lambda^{*})$ has just a single  peak for $\epsilon > 190 Jm^{-3}$ (Fig. \ref{Fig:HbSEnergy}). $R = 0$ and $R \rightarrow \infty$ correspond to the phases of the dispersed protofilaments and stable crystal structures, respectively.  As $\epsilon$ decreases below this, a local minimum develops in $\Omega(R;\Lambda^{*})$ whose position depends on $\epsilon$ and $\gamma$. The minimum critical bundle size $R_c$ occurs under the condition that $\Omega(R;\Lambda^{*})|_{R_c} = 0, \frac{\partial\Omega(R;\Lambda^{*})}{\partial R} |_{R_c} = 0$.  Combining the estimate $\epsilon \approx 38 Jm^{-3}$ for HbS \cite{Jones2003}, this yields a  value of $R_c  =  11nm$ and  $\gamma = 5.8  \mu Jm^{-2}$ ( Fig. \ref{Fig:HbSEnergy}), which are consistent with experimental observations \cite{Ferrone_03_01,Jones2003}.  A further reduction in $\epsilon$ leads to a decreasing value of $\Omega(R_c;\Lambda^{*})$ (Fig. \ref{Fig:HbSEnergy}).  We also note that the  torsional rigidity obtained by Turner et al is the same in our calculation, because we use their approximation for the elastic free energy;  this value for the rigidity is in agreement with experimental values. Finally, there always is an energy barrier for the transition from dispersed protofilaments to the metastable bundle phase, which is incorrectly predicted as a spontaneous process in reference \cite{Ferrone_03_01}. 

\section{Preliminary Results for a Brownian Dynamics Simulation of Bundle Formation}
\begin{figure}[h!]
\centering
\subfigure[]
{\includegraphics [width= 5.5cm]{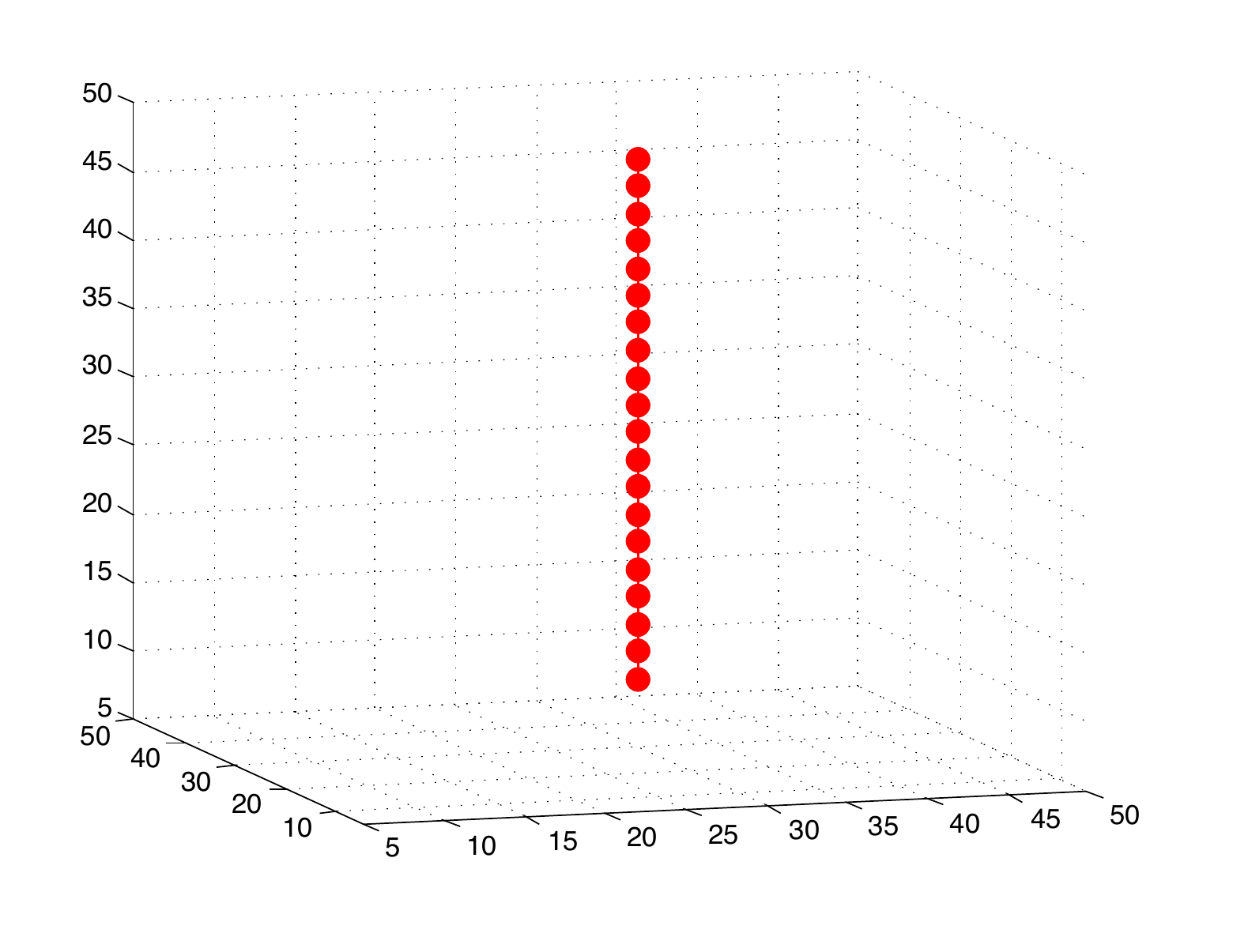}
\label{fig:SingleChainInit}
}
\subfigure[]{
\includegraphics [width= 5.5cm]{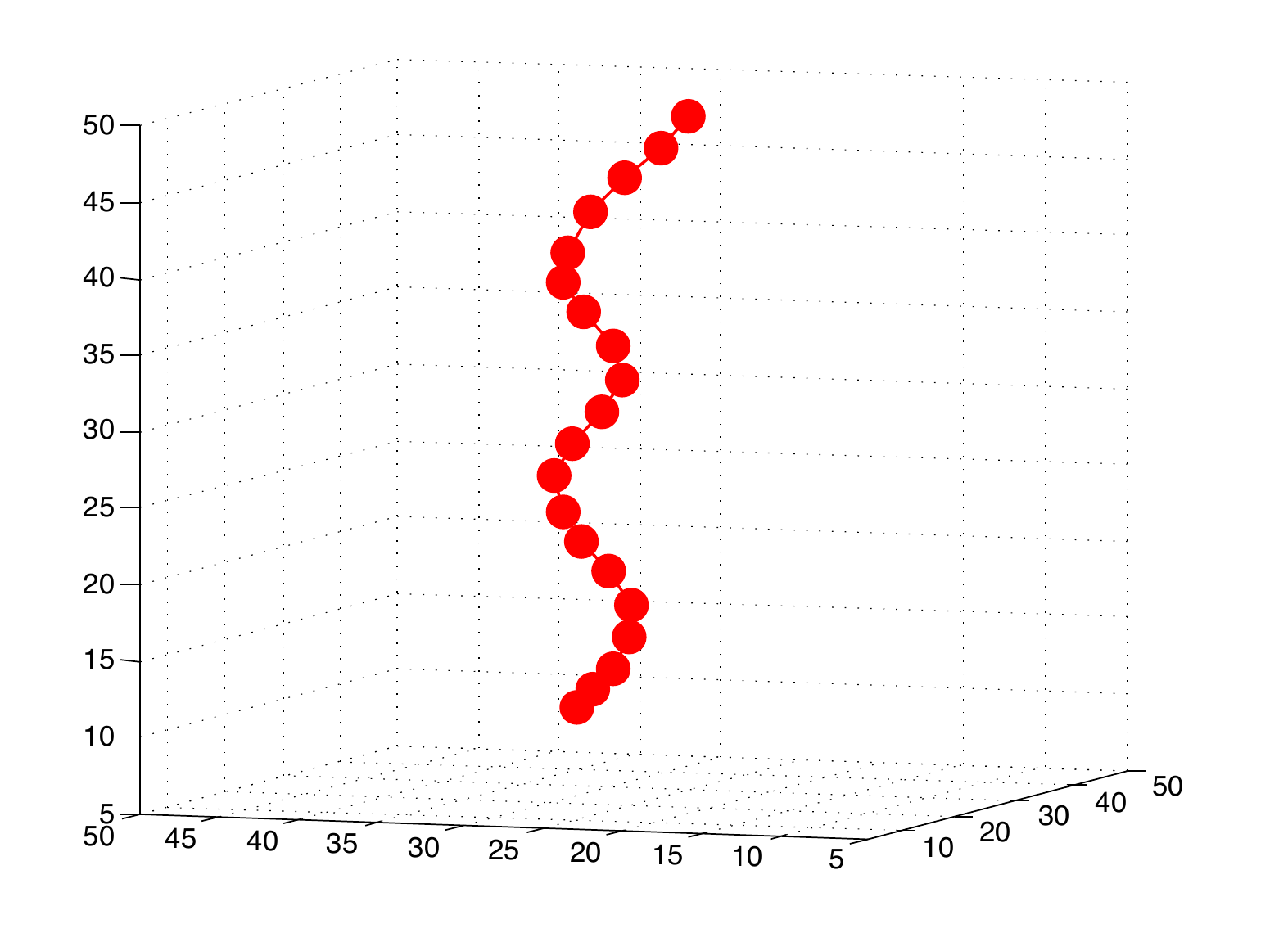}
\label{fig:SingleChainFinal}
}

\label{fig:SingleChain}
\caption[]{(a) a single straight chain with twenty monomers at t = 0, (b) a helical structure is formed at t = 10,000. }
\end{figure}

To understand and compare to experimental observations and our theoretical predictions of HbS, we are carrying out Brownian dynamics (BD) simulations  of bundle formation based on microscopic interactions.  In our simulation, the chiral filaments are described by the helical wormlike chain model \cite{Yamakawa1997},  in which the bending and twisting energy are incorporated into bead-spring polymers. The potential energy $U$  acting upon each monomer has three contributions:  the elastic energy associated with a single chain, the repulsive energy due to the excluded volume and the highly anisotropic short-range attractive energy between chains.
\begin{eqnarray}
U &=&  U_{chain} + U_{rep} + U_{c-c} \nonumber \\
U_{chain} &=& \frac{k_s}{2} \sum^{N}_{i=1}(r_{i,i-1} - l_0)^2 + \frac{\kappa_b}{2} \sum^{N}_{i=1}(\vec{u}_i - \vec{u}_{i-1} )^2 + \frac{\kappa_t}{2} \sum^{N}_{i=1}(\tau_i - \tau_0 )^2 \nonumber\\
U_{rep} &=& \sum^{N}_{i,j=1} 2\epsilon_{LJ} [(\frac{\sigma}{r_{ij}})^{12} - (\frac{\sigma}{r_{ij}})^{6}]
\label{Eq:BDpotential}
\end{eqnarray}
where $k_s$ is the spring constant, N the polymer length, $r_{i,i-1}$ the distance between two adjacent monomers ( ith and $(i-1)$th), $\kappa_b$ the bending stiffness, $\kappa_t$ the torsional stiffness, $\vec{u}_i $ the tangent vector  and $\tau_i$ the torsional angle on ith monomer.  $r_{ij}$ is the distance between two monomers, $\sigma$ the diameter of each monomer and $\epsilon_{LJ}$ the pair well depth.  The anisotropic attraction is modulated by the patchy model \cite{Noya_07_08}:
\begin{eqnarray}
U_{c-c} = \frac{1}{2} \sum^{N}_{i,j=1} U_{attr}(r_{ij})V_{ang}(\vec{r}_{ij}, \Omega_i, \Omega_j)
\label{Eq:Patchy}
\end{eqnarray}
where $U_{attr}$ is a Yukawa potential
\begin{eqnarray}
U_{attr}  = 
\begin{cases}
-A\sigma\frac{\exp(-Z(r_{ij}/\sigma - 1))}{r_{ij}}  &\mbox{if } r_{ij} \geq \sigma\\
0 &\mbox{if } r_{ij} < \sigma.
\end{cases}
\label{Eq:Yukawa}
\end{eqnarray}
Here $A$ is the energy strength and Z characterizes the range of attraction. 
\begin{eqnarray}
V_{ang}(\vec{r}_{ij}, \Omega_i, \Omega_j) = exp(-\frac{\theta_{k,ij}^2}{2\theta_0^2}) exp(-\frac{\theta_{l,ij}^2}{2\theta_0^2})
\label{Eq:Anglepart}
\end{eqnarray}
where $\theta_{k,ij}$ is the angle between patch k on the ith monomer and the interparticle displacement $\vec{r}_{ij}$. 
The particular pair of patches chosen (of the eight possible patches) is that which minimizes the magnitude of the angles $\theta_{k,ij}$ and $\theta_{l,ij}$ respectively. 
The parameter  $\theta_0$ is the standard deviation of the Gaussian distribution.  As an example, we apply the model for two simple cases: a single chiral chain and two binding  chiral chains. In the first case, a single straight chain is initially introduced in the system (Fig. \ref {fig:SingleChainInit}). As time evolves, a chain twists into a helical structure, shown in Fig. \ref {fig:SingleChainFinal}.  

\begin{figure}[h!]
\centering
\subfigure[]{
\includegraphics [width= 5.5cm]{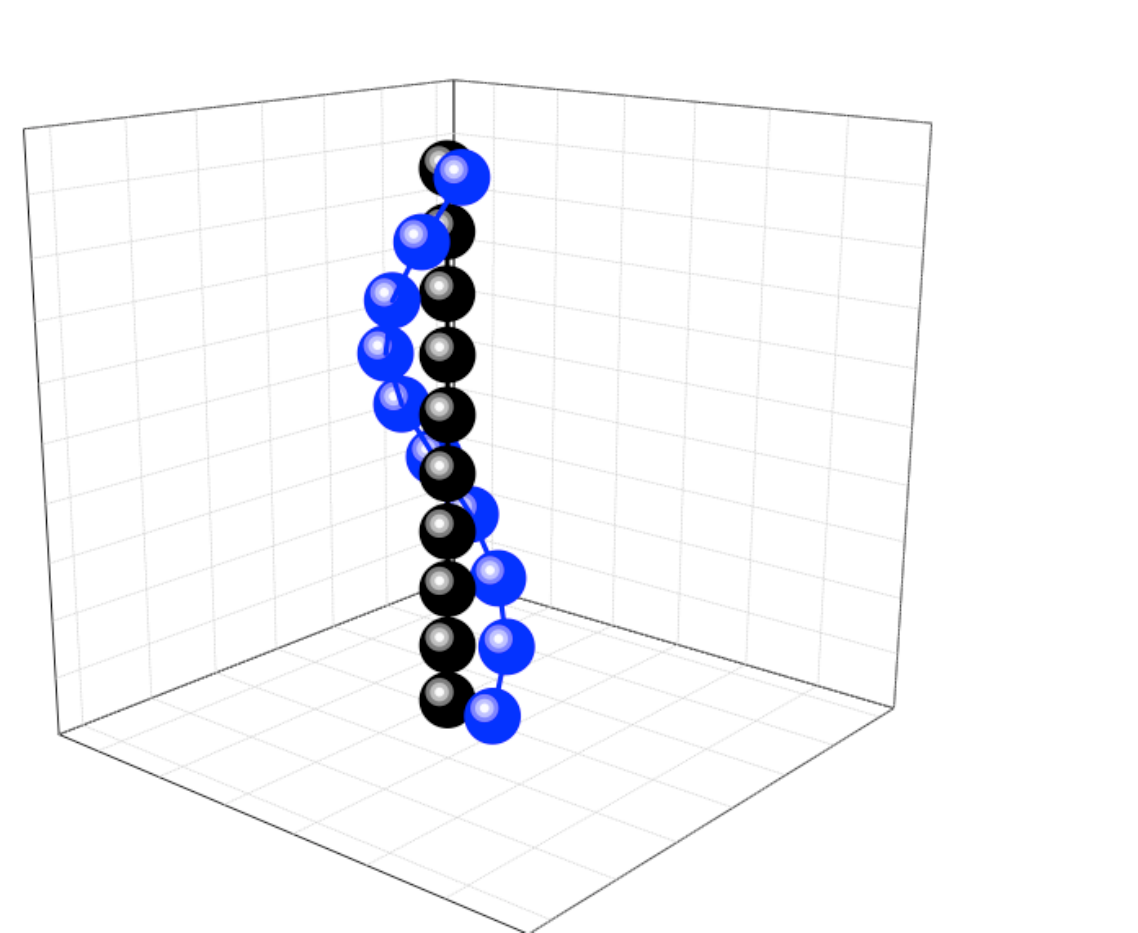}
\label{fig: ChainsInit}
}
\subfigure[]{
\includegraphics [width= 5.5cm]{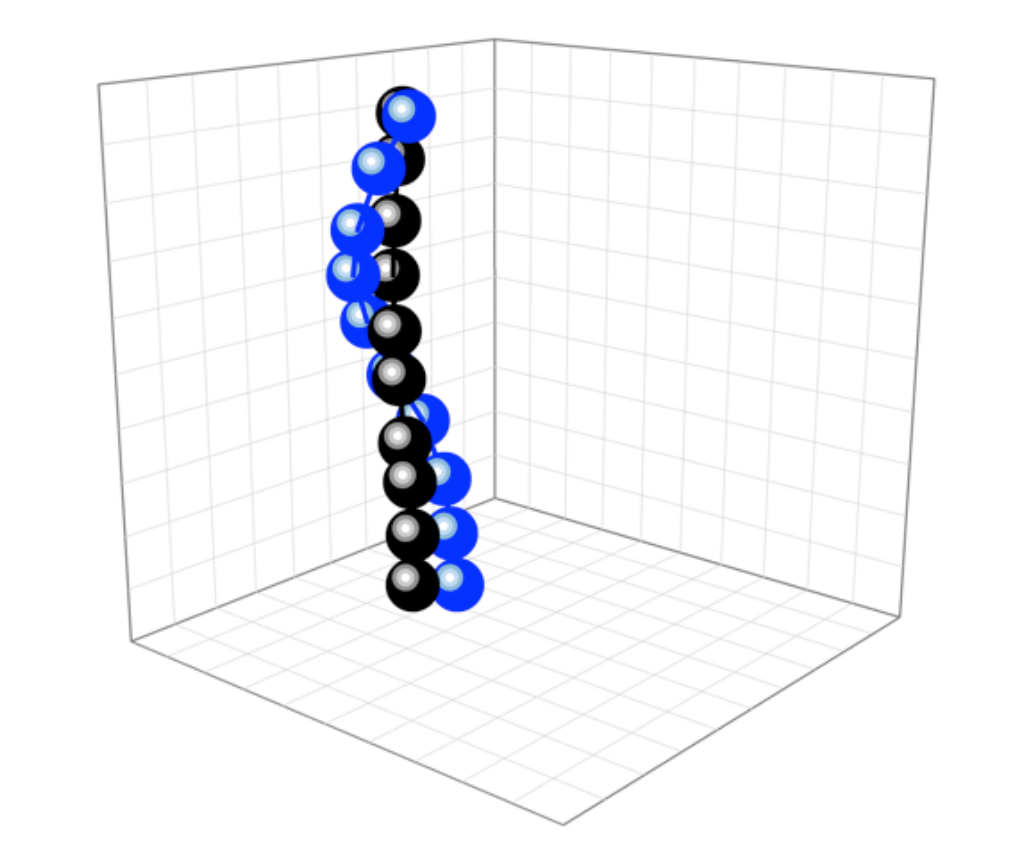}
\label{fig:ChainsInt}
}
\subfigure[]{
\includegraphics [width= 5.5cm]{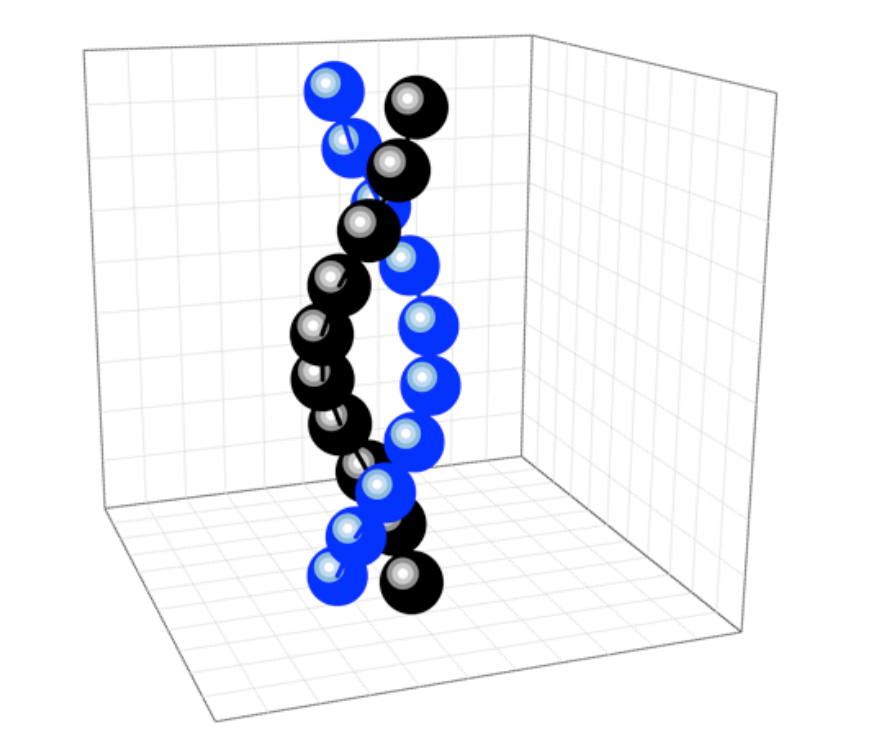}
\label{fig:ChainsFinal}
}
\label{fig:TwoChains}
\caption[]{(a) two bound chains with four equatorial patches (b) a configuration at t = 100, (c) at t = 5000, a double helical structure is formed.}
\end{figure}
In the second case of two binding chains, four patches are symmetrically arranged on the equator (perpendicular to the polymer bond) of each monomer. Two chains, one straight and one helical, are introduced, in which patches are aligned so that the patchy attraction is imposed at t = 0 (Fig. \ref {fig: ChainsInit}). Afterwards, two chains start intertwining and a configuration at t = 100 is shown in Fig.  \ref {fig:ChainsInt}.  
At t = 5000, two chains form a double helical structure (Fig. \ref {fig:ChainsFinal}).

\section{Kinetics of Cluster Formation in a 6-Patch Model of Protein Crystallization}

Many proteins are globular in shape but they have nonuniformly distributed surface charges that yield highly 
anisotropic interactions. In a recent work, the process of self-assembly in protein crystallization 
has been studied \cite{Liu2009}. In that work, they model spherical proteins using a patch model. 
The interaction between proteins takes into account an isotropic and a highly directional interaction.
The isotropic part was modeled as a square well of depth $\epsilon$ within a range of interparticle
separation between $\sigma$ and $\lambda\sigma$, $\sigma$ being the diameter of the particles and 
$\lambda=1.15$. For the anisotropic part, six patches uniformly distributed on the surface of each 
particle was considered. Then, the intrapatch interaction potential has a radial and an angular dependence:
\begin{eqnarray}
 u_p(r,\Omega_i,\Omega_j)= u^{SW}(r)f(\Omega_i,\Omega_j).
\end{eqnarray}
where $u^{SW}$ is defined in the same way as the isotropic part, but in this case with a deeper 
well depth $\epsilon_p=5\epsilon$ and narrow well width $\lambda_p=1.05$. The angular part was 
defined as:
\begin{center}
$
 f(\Omega_i,\Omega_j)=\left\lbrace \begin{array}{lr}
1& \theta_i, \theta_j \le \delta \\
0& otherwise
\end{array}
\right.
$\\
\end{center}
$\theta_i$ being the angle between the patch orientation, determined by a unit normal vector 
centered in the patch, and the line that connect the center of mass of the particles. $\delta=0.259$ 
determines the patch size.

\begin{figure}[h!]
 \centering
 \includegraphics[width=0.6\linewidth]{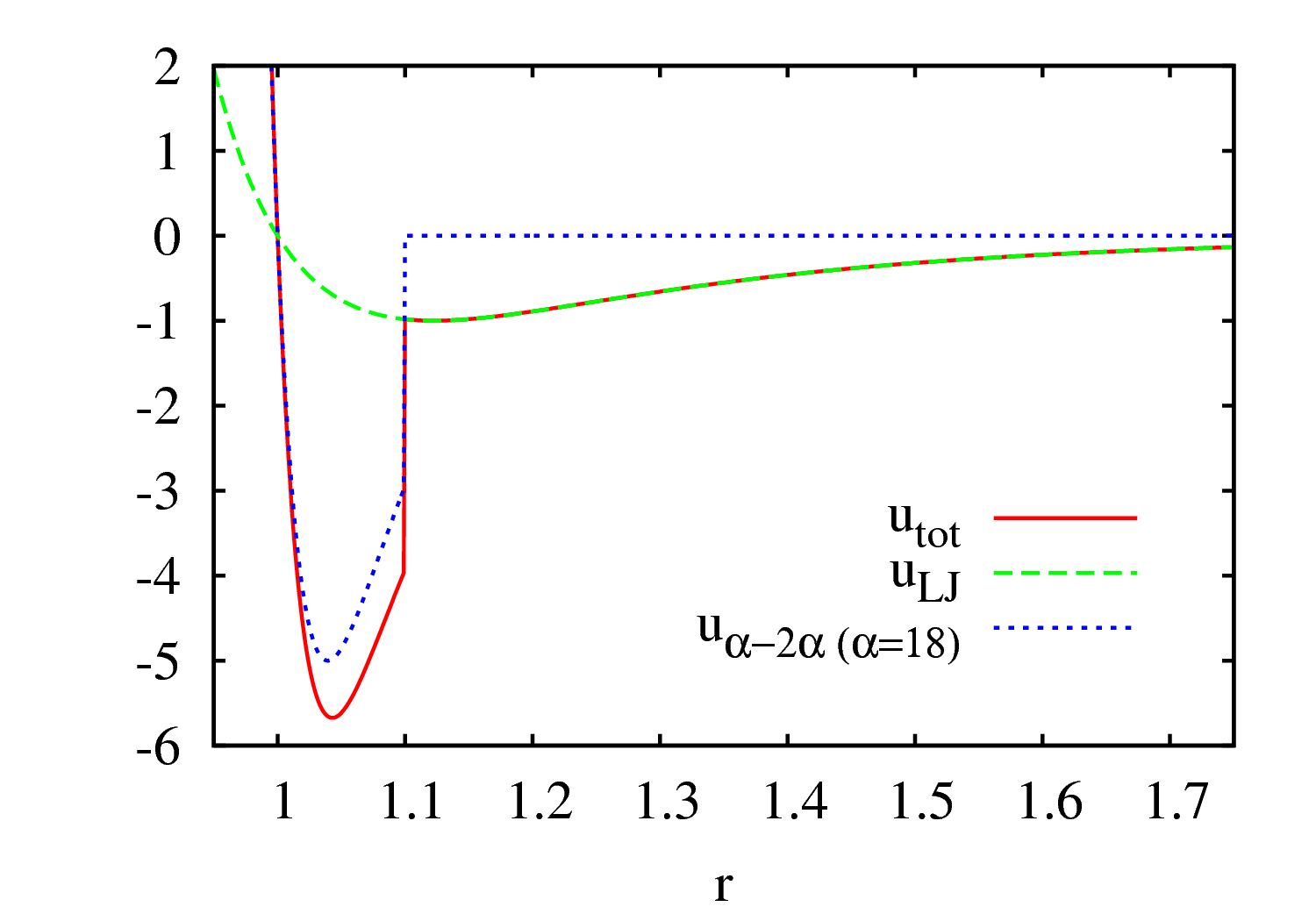}
 \caption{Plot of the proposed potential. The isotropic part of the potential is formed by the 
Lennard-Jones potential (dashed line). The patch-patch interaction is modeled by the $\alpha-2\alpha$ 
potential with $\alpha=18$ and a cut-off of $r_c=1.1$ (dotted line). When the patches in two 
different particles are aligned, and the distance between the particles is less than $r_c$, the total 
potential resulting is the combination of the isotropic and the directional parts (solid line).}
 \label{fig: potential_a2a}
\end{figure}

In our work, we are interested in the dynamics of the nucleation and growth process of a
six-patch model of globular protein using Brownian dynamics simulations. In particular we study 
the kinetics of the process by which chains form and develop local crystal order that then optimizes the
crystal nucleation process.
\begin{figure}[h!]
 \centering
 \includegraphics[width=0.6\linewidth]{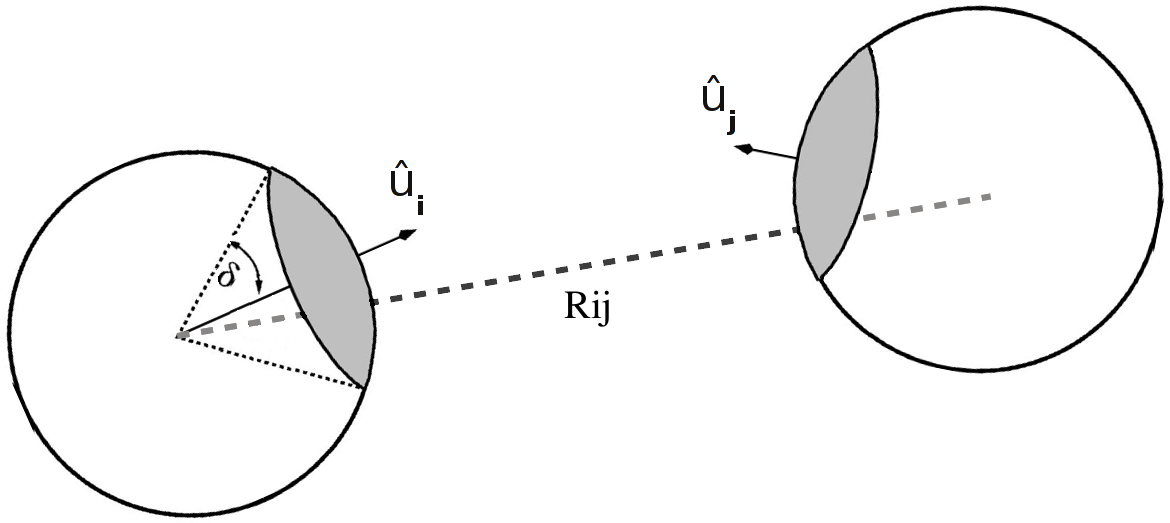}
 \caption{Sketch of patchy particles. A patch is defined by a solid angle with 
half opening angle $\delta$ about an axis $\hat{u}_i$. Two particles attract 
each other if the angle between two patches in their surfaces is within a given range.}
 \label{fig: patch_scheme}
\end{figure}\\
Our model consists in a continuous version of the previous 6-patch model. Figure \ref{fig: patch_scheme} shows a sketch of the patchy particles. The interparticle 
potential is composed of an isotropic interaction and a directional dependent interaction:
\begin{eqnarray}
 u(r,\Omega)= \epsilon u^{LJ}(r)+\epsilon_p u^{\alpha-2\alpha}(r)f(\Omega)
\end{eqnarray}
$u^{LJ}$ is the Lennard-Jones potential and the $u^{\alpha-2\alpha}(r)$ potential is 
\begin{eqnarray}
 u^{\alpha-2\alpha}(r)=4 \left [ \left ( \frac{\sigma}{r}  \right )^{2\alpha}-\left ( \frac{\sigma}{r}  \right )^{\alpha}  \right ].
\end{eqnarray}
The angular modulation of the interaction taking into account the alignment of the patches is given by\\
\begin{center}
$
 f(\Omega)=\left\lbrace \begin{array}{lr}
1&\theta_{i},\theta_{j}\le \delta\\
0& otherwise
\end{array}
\right.
$\\
\end{center}
where $\theta_{i}$ is the angle between any patch and the line joining the center of mass 
of the particles and $\delta$ is the patch half opening angle defining the size of the patch. 
Figure \ref{fig: potential_a2a} shows the potential we are considering in this work. 

\begin{figure}[h!]
 (a)
 \begin{center}
  \includegraphics[width=0.6\linewidth]{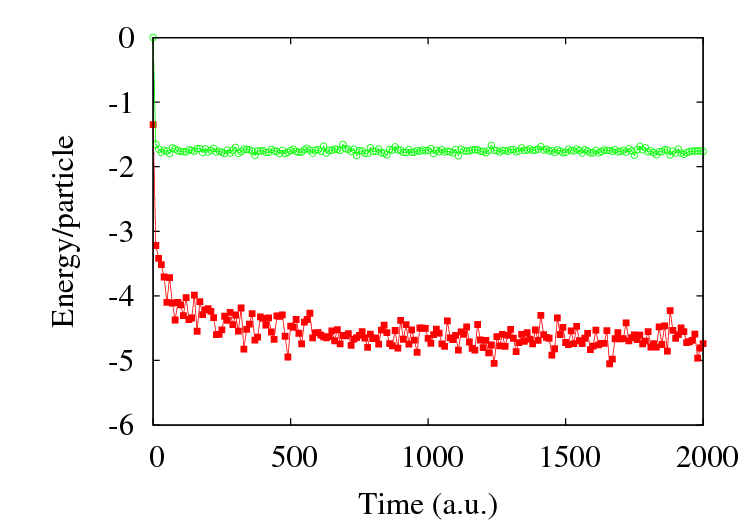}          
 \end{center}
 (b)
 \begin{center}
  \includegraphics[width=0.475\linewidth]{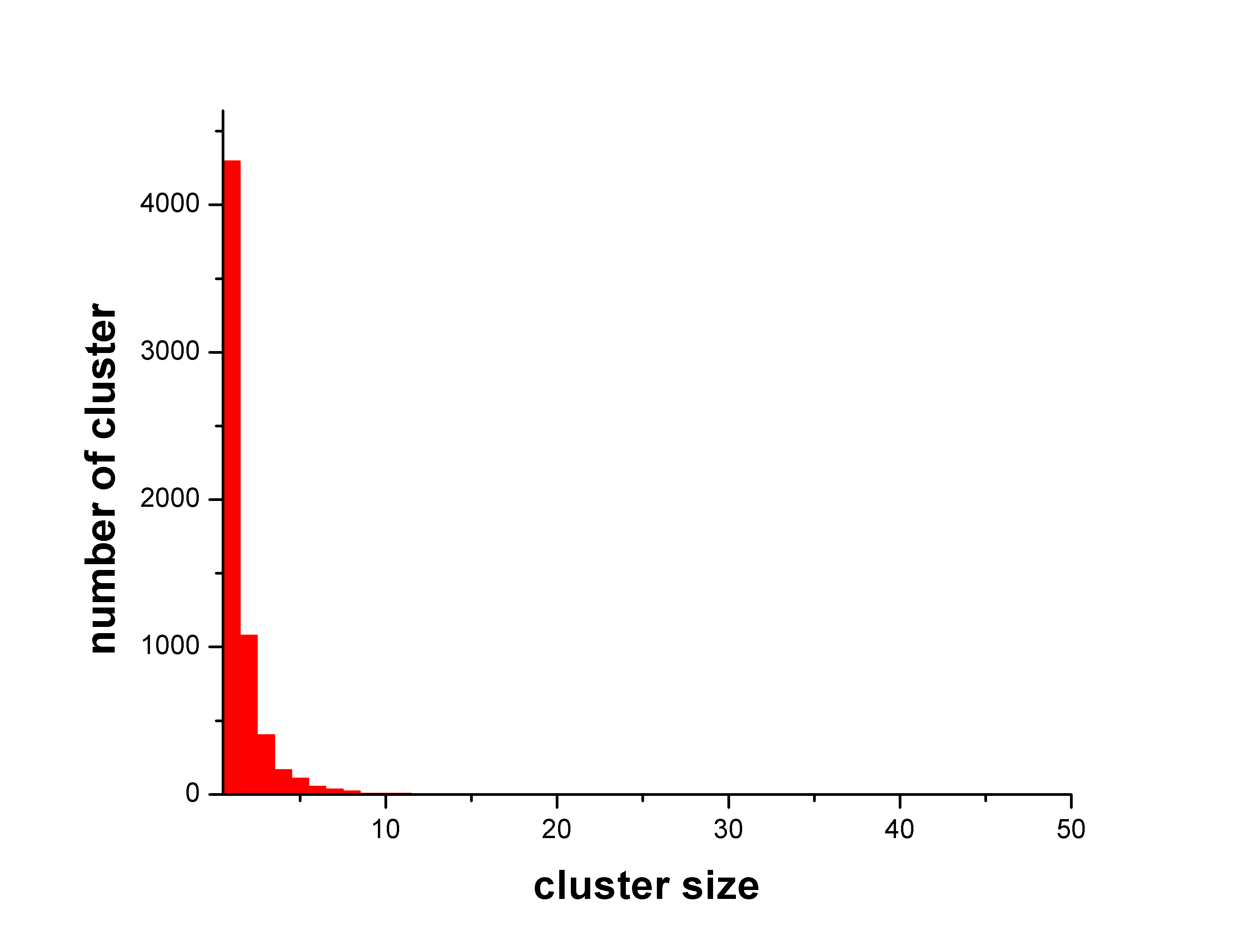}
  \includegraphics[width=0.475\linewidth]{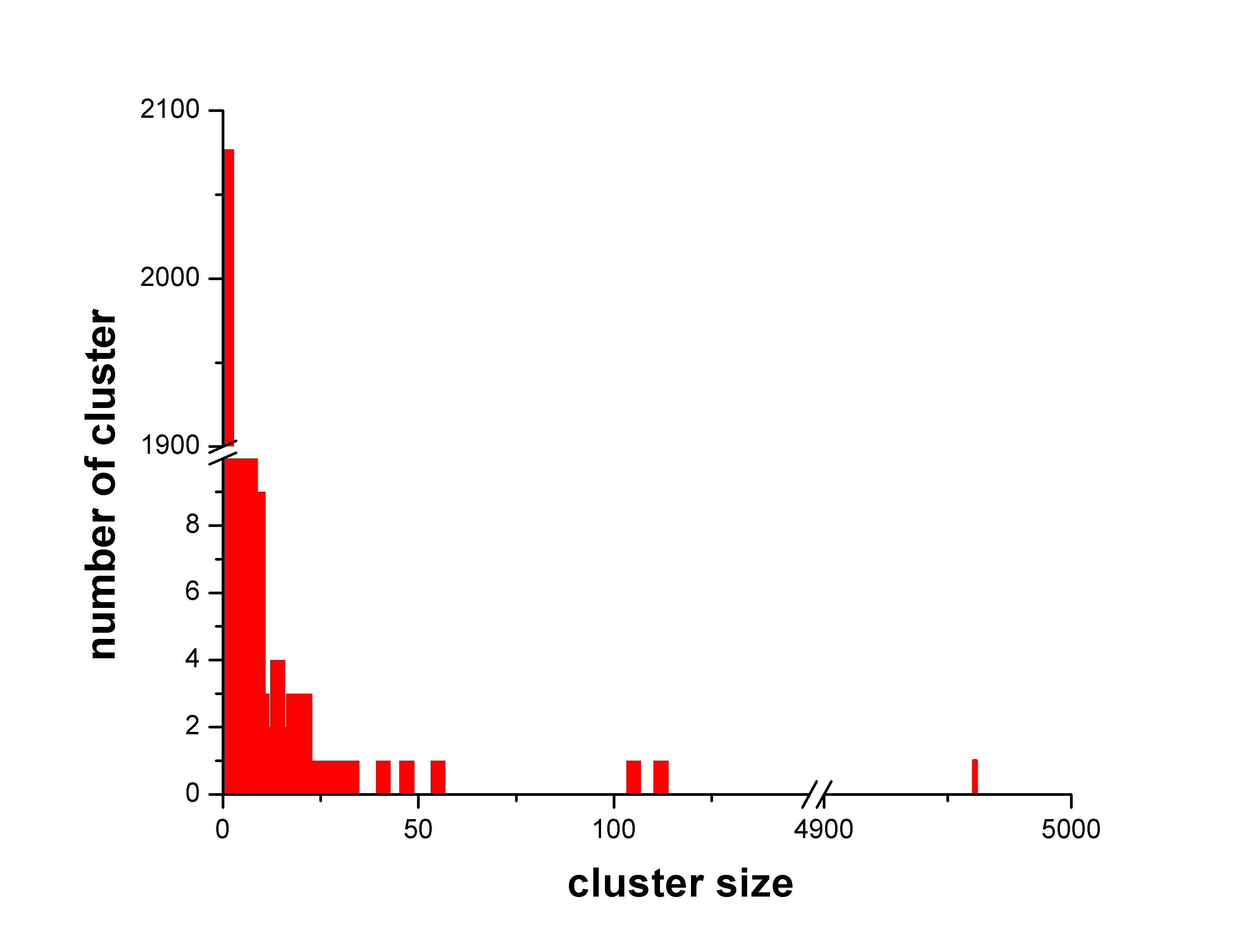}
 \end{center}
 \caption{(a) Energy per particle for two different ranges of the isotropic part of the potential. In the short range case (open circles), 
the energy per particle slightly fluctuates around the value $u/k_BT\sim-1.75$ while in the long range case (filled squares) the energy exhibit larger fluctuations around a lower value. 
(b) Cluster distributions for the two previous conditions. The long range 
potential (right panel) induces the formation of a big cluster from the early stages of the simulation. On the contrary, for the short range potential 
(left panel), we only observe a small size cluster distribution.}
 \label{fig: potential_ranges}
\end{figure}

In order to characterize the dynamics of the cluster formation, we will determine how $S(q,t)$ changes with time. 
We will also interested in characterizing the structure of the  clusters that form by calculating the local bond-order 
parameter $q_6(i)$. The quantity is a convenient measure of the local crystal order.
\newpage
A preliminary result indicates that the range of the isotropic part of the potential plays an important role in the initial 
states of cluster formation. We observed from simulation that the long range isotropic potential ($r_{cut}=2.5$) enhances and induces the particles to 
aggregate in a short time when compared with the short range isotropic potential ($r_{cut}=1.15$). 
Figure \ref{fig: potential_ranges} shows the energy per particle and the cluster 
distribution for both the long and short range isotropic parts of the potential.

\newpage
\newpage
\section*{Acknowledgements} This work is supported by grants from the National Science Foundation (DMR- 0702890) and the G. Harold and Leila Y. Mathers Foundation. SJK and AC are supported by NSF NIRT grant CTS0609318.

\bibliographystyle{elsarticle-num}
\bibliography{biblio}

\begin{thebibliography}{10}
\expandafter\ifx\csname url\endcsname\relax
  \def\url#1{\texttt{#1}}\fi
\expandafter\ifx\csname urlprefix\endcsname\relax\def\urlprefix{URL }\fi
\expandafter\ifx\csname href\endcsname\relax
  \def\href#1#2{#2} \def\path#1{#1}\fi

\bibitem{Bromberg_05_01}
L.~Bromberg, J.~Rashba-Step, T.~Scott, Insulin particle formation in
  supersaturated aqueous solutions of poly(ethylene glycol), Biophys. Jour. 89
  (2005) 3424--3433.

\bibitem{Grason_07_01}
G.~M. Grason, R.~F. Bruinsma, Phy. Rev. Lett. 99 (2007) 098101.

\bibitem{Gunton_07_01}
J.~D. Gunton, A.~Shiryayev, D.~L. Pagan, Protein Condensation: Kinetic Pathways
  to Crystallization and Disease, Cambridge University Press, 2007.

\bibitem{Ferrone_03_01}
M.~S. Turner, R.~W. Briehl, F.~A. Ferrone, R.~Josephs, Twisted protein
  aggregates and disease: the stability of sickle hemoglobin fibers, Phys. Rev.
  Lett. 90 (2003) 128103.

\bibitem{Oxtoby1992}
D.~W. Oxtoby, J. Phys.: Condens. Matter 4 (1992) 7627.

\bibitem{Onuki_02_01}
A.~Onuki, Phase Transition Dynamics, Cambridge University Press, Cambridge, UK,
  2002.

\bibitem{Jones2003}
C.~W. Jones, J.~C. Wang, F.~A. Ferrone, R.~W. Briehland, M.~S. Turner, Faraday
  Discuss. 123 (2003) 221.

\bibitem{Yamakawa1997}
H.~Yamakawa, Helical Wormlike Chains in Polymer Solutions, Springer, 1997.

\bibitem{Noya_07_08}
E.G.Noya, C.Vega, J.~P.~K. Doye, A.~A. Louis, J. Chem. Phys. 127 (2007) 054501.

\bibitem{Liu2009}
H.~Liu, S.~Kumar, J.~Douglas, Self-assembly-induced protein crystallization,
  Phys. Rev. Let. 103 (2009) 018101.

\end{thebibliography}







\end{document}